\documentclass[amsmath,twocolumn,pra]{revtex4-1}
\usepackage{graphicx}
\usepackage{times}
\usepackage{epsfig}
\usepackage{multirow}
\usepackage{dcolumn}
\usepackage{booktabs}
\usepackage{revsymb}
\usepackage{mathrsfs}
\begin{document}

\title{Relativistic corrections to the thermal interaction of bound particles}

\author{D. Solovyev}
\email{d.solovyev@spbu.ru}
\affiliation{Department of Physics, St.Petersburg State University, St.Petersburg, 198504, Russia}

 \author{T. Zalialiutdinov}
\affiliation{Department of Physics, St.Petersburg State University, St.Petersburg, 198504, Russia}
 \author{A. Anikin}
\affiliation{Department of Physics, St.Petersburg State University, St.Petersburg, 198504, Russia}

\begin{abstract}
This paper discusses relativistic corrections to the thermal Coulomb potential for simple atomic systems. The theoretical description of the revealed thermal corrections is carried out within the framework of relativistic quantum electrodymamics (QED). As a result, thermal corrections to the fine and hyperfine strucutres of atomic levels are introduced. The theory presented in this paper is based on the assumption that the atom is placed in a thermal environment created by the blackbody radiation (BBR). The numerical results allow us to expect hteir significance for modern experiments and testing the fundamental interactions.
\end{abstract}
\maketitle

\section{Introduction}

Since the early days of quantum mechanics (QM), the study of characteristics of atoms, such as the energy of bound states, has played a key role in the development of modern quantum field theory and its practical application in various fields of physics. Further development of quantum mechanics led to the creation of a quantum electrodynamical description (QED) of atoms and the evaluation of the corresponding relativistic QED corrections to bound energies \citep{Bethe,Akhiezer,Sob,Berest,lindgren,LabKlim,Greiner}. Subsequent experimental observations and their growing accuracy have required taking into accounting more complex effects, such as, for example, a multitude of radiative QED corrections mutually providing a versatile verification of fundamental physics. Much effort has been put into this type of theoretical research, see, for example, \cite{Andr,shabaev,Indelicato_2019}.

Nowadays, the most accurate atomic experiments can be attributed to three directions corresponding to measurements of the transition frequencies in hydrogen \cite{Parthey,Mat} with a relative uncertainty of $4.2\times 10^{-15}$, in helium \cite{Roo,Zheng}, where the experimental accuracy reaches the level of several parts in $10^{-12}$, and in atomic clocks possessing an accuracy about $10^{-17}$ for time scaling \cite{AtCl-Cs,AtCl-Sr}. Such precise experiments required theoretical calculations of various QED effects at the $\alpha^6 m^2/M$ and $\alpha^7m$ levels, see \cite{Mohr-2016} and references therein, where $\alpha$ is the fine structure constant, $m$ and $M$ are the electron and nuclear masses, respectively. Apart from accurate theoretical calculations of the binding energies in the hydrogen atom, the fine structure, and the isotope shift of the low-lying states of helium tend to serve as an independent tool for testing of fundamental interactions. Similar to well-studied one-electron atomic systems the measured transition frequencies should be compared with the theoretical calculations pursuing the search of possible discrepancy \cite{PPY-2017}.

Such extraordinary calculations however pay attention to the effects of the other type: corrections induced by the external blackbody radiation (BBR) field. The influence of the BBR field is well-known in atomic physics, it is manifested in the existence of a Stark shift of bound states. The theory and corresponding calculations for one- and few-electron atoms were presented in \cite{farley} in the framework of the QM approach. These calculations were continued for the case of atomic clocks (many-particle systems) in \cite{Saf,SKC,Porsev} and are the subject of theoretical investigations in present days. Not long ago, the QED derivation of the Stark shift induced by the BBR field was performed in \cite{SLP-QED} and, subsequently, applied to calculations in the helium atom \cite{jphysb2017}.

Recently in \cite{S-TQED} a QED description of the effects induced by the BBR field for one-electron atomic systems was presented. Although this theory is based on Thermal Quantum Electrodynamics (TQED) pioneered in \cite{Dol,Don,DHR}, thermal Coulomb interaction was first rigorously introduced in \cite{S-TQED}. In particular, it was found that the energy shift associated to this interaction can exceed the corresponding Stark shift. The effect of thermal one-photon exchange between the bound electrons and nucleus was investigated in \cite{SZA-2020} for the helium atom. From the results obtained in this work it follows that in two-electron atomic system the thermal shift reaches the level of the experimental accuracy \cite{KS-Hessel} at room temperature. 

In view of the close attention to the verification of fundamental physical interactions in such experiments and the search for new constraints on dark matter \cite{Kennedy}, the derivation of thermal effects leading to fine and hyperfine splitting of levels is of considerable interest. This problem can be solved using the formalism presented in \cite{S-TQED}. Then, the representation of the thermal photon propagator in the thermal Coulomb gauge admits an explicit analogy with the case of zero vacuum, see \cite{Berest}, i.e. arising relativistic corrections are easily extended to thermal ones. It can be expected that thermal corrections of these types can serve for further testing of fundamental interactions on atomic systems.

In this paper, the relativistic thermal corrections arising from the scalar and transversal part of the thermal photon propagator are evaluated. Then their contributions for one- and two-electron atoms are estimated. All the derivations are performed within the framework of the rigorous quantum electrodynamics at finite temperatures and are applicable to H-like ions. For clarity, the mass of particles and speed of light $c$ are written out explicitly in the basic formulas.

\section{Thermal Coulomb interaction: relativistic corrections}

Starting with the description of the interaction of two charges, one can use the relation from textbooks (see, for example, \cite{Akhiezer}) connecting the nuclear current, $j^\nu(x')$, with the field, $A_\mu(x)$, it creates:
\begin{eqnarray}
\label{1}
A_\mu(x) = \int d^4x D_{\mu\nu}(x,x')j^\nu(x'),
\end{eqnarray}
where $x=(t,\vec{r})$ represents the four-dimensional coordinate vector ($t$ represents time and $\vec{r}$ denotes a space vector), $D_{\mu\nu}(x,x')$ is the Green's function of the photon, and $\mu$, $\nu$ are the indices running the values $0,1,2,3$. Then, the zero component of $A_\mu(x)$ corresponds to the Coulomb interaction, and the components $1,2,3$ are the transversal part, which gives the interaction of retardation and advance.
According to \cite{Dol,Don,DHR}, the photon Green's function (photon propagator) is represented by the sum of two contributions, which are the result of expectation value on the states of zero and heated vacuum, $D_{\mu\nu}(x,x') = D_{\mu\nu}^0(x,x')+D^\beta_{\mu\nu}(x,x')$, respectively.

\begin{figure}[hbtp]
	\centering
	\includegraphics[scale=0.125]{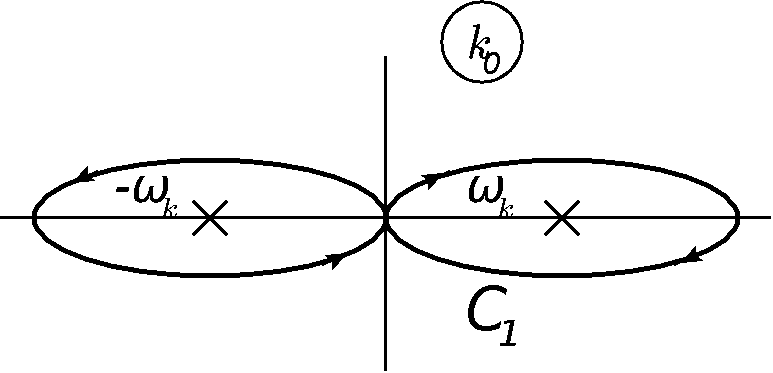} 
	\caption{Integration contour $C_1$ in $k_0$ plane of Eq. (\ref{2}).Arrows on the contour define the pole-bypass rule. The poles $\pm\omega_k$ are denoted with $\times$ marks.}
	\label{Fig-1}
\end{figure}

Thermal interaction can be introduced by analogy, see \cite{S-TQED}, when the 'ordinary' photon Green's function is replaced by the thermal one, $D_{\mu\nu}^\beta(x_1,x_2)$, \cite{Dol,Don,DHR}. In \cite{S-TQED} is was established that the thermal part of photon propagator $D_{\mu\nu}^\beta(x,x')$ is given by the Hadamard propagation function \cite{Akhiezer,Greiner} and, therefore, admits a different (equivalent) form:
\begin{eqnarray}
\label{2}
D_{\mu \nu}^{\beta}(x, x') =
- 4\pi g_{\mu\nu}\int\limits_{C_1}\frac{d^4k}{(2\pi)^4} \frac{e^{ik(x-x')}}{k^2}n_\beta(|\vec{k}|),
\end{eqnarray}
where $g_{\mu\, \nu}$ is the metric tensor, $k^2=k_0^2-\vec{k}^2$ and $n_{\beta}$ is the Planck's distribution function. The contour of integration in $k_0$-plane for Eq. (\ref{2}) is given in Fig.~\ref{Fig-1}.

Thermal photon propagator in the form Eq. (\ref{2}) has the advantage of allowing the introduction of gauges in complete analogy with the 'ordinary' QED theory, see \cite{S-TQED}. Then in the Coulomb gauge the function $D_{\mu\nu}^\beta(x_1,x_2)$ recasts into
\begin{eqnarray}
\label{3}
D_{00}^{\beta}(x, x') &=&  4\pi i \int\limits_{C_1}\frac{d^4k}{(2\pi)^4}\frac{e^{i k (x-x')}}{\vec{k}^2}n_\beta(\omega),\qquad
\\
\nonumber
D_{ij}^{\beta}(x, x') &=& 4\pi i \int\limits_{C_1}\frac{d^4k}{(2\pi)^4}\frac{e^{i k (x-x')}}{k^2}n_\beta(\omega)\left(\delta_{ij}-\frac{k_i k_j}{\vec{k}^2}\right).
\end{eqnarray}

Concentrating first on the thermal Coulomb interaction, we consider the zero component of the thermal photon propagator, $D_{00}^{\beta}(x, x')$. Substituting this into Eq. (\ref{1}) the thermal potential for the point-like nucleus in the static limit can be found.
For the conciseness we omit the discussion and employ the procedure proposed in \cite{S-TQED}, where the appropriate analytical calculations of the integral over $\kappa$ was obtained as
\begin{eqnarray}
\label{4}
V^\beta(r)=-\frac{4e^2}{\pi}
\left(-\frac{\gamma}{\beta}+\frac{i }{2 r}\ln \left[\frac{\Gamma \left(1+\frac{i r}{\beta}\right)}{\Gamma \left(1-\frac{i r}{\beta}\right)}\right]\right),
\end{eqnarray}
where $\beta\equiv 1/(k_B T)$ ($k_B$ is the Boltzmann constant and $T$ is the temperature in kelvin), $r$ is the modulus (length) of the corresponding radius vector for the interpartcle distance, $\Gamma$ is the gamma function and $\gamma$ is the Euler-Mascheroni constant, $\gamma\simeq 0.577216$.

The potential (\ref{4}) was used to find thermal corrections to the energy of a bound electron in the hydrogen atom \cite{S-TQED} and helium \cite{SZA-2020}, where the thermal correction tuned out to be of the order of the experimental accuracy \cite{KS-Hessel}.
As it should be, in the lowest order the heat bath environment removes the orbital momentum degeneracy. To find other corrections (to the fine and hyperfine structure), one should turn to the Pauli approximation or determine the relativistic corrections proportional to $1/c^2$, where $c$ is the speed of light, see \cite{Akhiezer,Greiner,LabKlim,Berest}. Within the second order approximation, the particle interaction operator (in the case of zero vacuum) in momentum representation and Coulomb gauge has the form \cite{Berest}:
\begin{eqnarray}
\label{5}
U(\vec{p}_1,\vec{p}_2,\vec{k}) = 4\pi e^2\left[\frac{1}{\vec{k}^2}-\frac{1}{8m_1^2c^2}-\frac{1}{8m_2^2c^2} 
\right.
\nonumber
\\
+
\left.
\frac{(\vec{k}\vec{p}_1)(\vec{k}\vec{p}_2)}{m_1 m_2 c^2\vec{k}^4} - \frac{\vec{p}_1\vec{p}_2}{m_1 m_2 c^2\vec{k}^2}  + \frac{i\vec{\sigma}_1[\vec{k}\times\vec{p}_1]}{4m_1^2 c^2 \vec{k}^2}-\frac{i\vec{\sigma}_2[\vec{k}\times\vec{p}_2]}{4m_2^2 c^2 \vec{k}^2} 
\right.
\\
\left.
\nonumber
-
\frac{i\vec{\sigma}_1[\vec{k}\times\vec{p}_2]}{2m_1 m_2 c^2 \vec{k}^2} + \frac{i\vec{\sigma}_2[\vec{k}\times\vec{p}_1]}{2m_1 m_2 c^2 \vec{k}^2} + \frac{(\vec{\sigma}_1\vec{k})(\vec{\sigma}_2\vec{k})}{4m_1 m_2 c^2\vec{k}^2}-\frac{\vec{\sigma}_1\vec{\sigma}_2}{4m_1 m_2 c^2}
\right].
\end{eqnarray}
Here $\vec{p}$ is the electron momentum operator, $\vec{\sigma}$ is the Pauli matrix, $m$ is the particle mass and the index 1 or 2 refers to the corresponding particle.

The Fourier component Eq. (\ref{5}) contains relativistic corrections arising from the scalar and transverse parts of the photon propagator $D_{00}(x,x')$ and $D_{ij}(x,x')$, respectively. The latter also corresponds to the Breit-Pauli interaction. Applying the Fourier transform, $\int\frac{d^3k}{(2\pi)^3}e^{i\vec{k}\vec{r}}$, to the scattering amplitude $U(\vec{p}_1,\vec{p}_2,\vec{k})$, the coordinate representation can be found. Then the operators $\vec{p}_1$ and $\vec{p}_2$ should be replaced by the $\vec{p}_1 = -i\nabla_1$ and $\vec{p}_2 = -i\nabla_2$. In the thermal case, however, the Fourier transform to coordinate representation is given as $2\int\frac{d^3k}{(2\pi)^3}n_\beta(|\vec{k}|)e^{i\vec{k}\vec{r}}$, where the factor 2 occurs in a result of integration along the contour  $C_1$.

The result of such a Fourier transform for the first term in Eq. (\ref{5}) results in the expression (\ref{4}), where the regularization of divergent contribution at $|\vec{k}|\rightarrow 0$ was performed by introducing a coincidence limit \cite{S-TQED}. The same can be easily given for the second and third terms in Eq. (\ref{5}). After the integration over angles one can find
\begin{eqnarray}
\label{6}
(2)+(3)\rightarrow - \frac{e^2}{2\pi}\left[\frac{1}{m_1^2 c^2}+\frac{1}{m_2^2 c^2}\right]\int\limits_0^\infty d\kappa\, n_\beta(\kappa)\frac{\kappa\sin\kappa r_{12}}{r_{12}}.
\end{eqnarray}
Then the series expansion in small values of $r_{12}$ reveals a constant contribution proportional to $\int\limits_0^\infty d\kappa\,\kappa^2n_\beta(\kappa)$. In principle, this contribution is state independent and, therefore, vanishes for the difference between the energy states of the atom. But, it can be found immediately that the coincidence limit \cite{S-TQED} regularizes such contributions along with divergences. In other words, subtraction of the limit $r_{12}\rightarrow 0$ in Eq. (\ref{6}) gives the regular expression, which is
\begin{eqnarray}
\label{7}
(2)+(3)\rightarrow \frac{2e^2}{\pi}\frac{\zeta(5)}{\beta^5}r_{12}^2\left[\frac{1}{m_1^2 c^2}+\frac{1}{m_2^2 c^2}\right].
\end{eqnarray}
Here we have integrated over $\kappa\equiv |\vec{k}|$ and $\zeta(s)$ gives the Riemann zeta function.

In the fourth term in Eq. (\ref{5}) first note that $\vec{k}$ in the numerator can be obtained by the gradient action on the exponent:
\begin{eqnarray}
\label{8}
\frac{4\pi e^2}{m_1 m_2 c^2}2\int\frac{d^3k}{(2\pi)^3}n_\beta(\kappa)e^{i\vec{k}\vec{r}_{12}}\frac{(\vec{k}\vec{p}_1)(\vec{k}\vec{p}_2)}{\vec{k}^4} = 
\\
\nonumber
\frac{4\pi e^2}{m_1 m_2 c^2}2\int\frac{d^3k}{(2\pi)^3}n_\beta(\kappa)\frac{(\vec{\nabla}_1\vec{p}_1)(\vec{\nabla}_2\vec{p}_2)}{\vec{k}^4}e^{i\vec{k}\vec{r}_{12}}.
\end{eqnarray}
Then, integrating over angles in the expression (\ref{8}) and acting by the gradient operators, the formula (\ref{8}) reduces to
\begin{eqnarray}
\label{9}
\frac{4e^2}{\pi m_1 m_2 c^2}\int\limits_0^\infty d\kappa\, \frac{n_\beta(\kappa)}{\kappa^2}\left[\frac{\cos \kappa r_{12}}{r_{12}^2}-\frac{\sin\kappa r_{12}}{\kappa r_{12}^3}\right](\vec{p}_1\vec{p}_2)
\\
\nonumber
-\frac{4e^2}{\pi m_1 m_2 c^2}\int\limits_0^\infty d\kappa\, \frac{n_\beta(\kappa)}{\kappa^2}\left[\frac{3\cos \kappa r_{12}}{r_{12}^4}+\frac{3\sin\kappa r_{12}}{\kappa r_{12}^5}
\right.
\\
\nonumber
\left.
-\frac{\kappa\sin\kappa r_{12}}{r_{12}^3}\right](\vec{r}_{12}\vec{p}_1)(\vec{r}_{12}\vec{p}_2)
.
\end{eqnarray}
The coincidence limit $r_{12}\rightarrow 0$ in this case is
\begin{eqnarray}
\label{10}
-\frac{4e^2}{3\pi m_1 m_2 c^2}\int\limits_0^\infty d\kappa\,n_\beta(\kappa),
\end{eqnarray}
which cancels the divergence in Eq. (\ref{9}). Finally, the Fourier transform of fourth term in Eq. (\ref{5}) reduces to
\begin{eqnarray}
\label{11}
4\pi e^2\frac{(\vec{k}\vec{p}_1)(\vec{k}\vec{p}_2)}{m_1 m_2 c^2\vec{k}^4} \rightarrow \frac{4\zeta(3) e^2}{15\pi \beta^3 m_1 m_2 c^2}\times
\\
\nonumber
\left[r_{12}^2(\vec{p}_1\vec{p}_2) + 2(\vec{r}_{12}\vec{p}_1)(\vec{r}_{12}\vec{p}_2)\right].
\end{eqnarray}

Evaluation of the fifth contribution in Eq. (\ref{5}) repeats the calculation of the first one. The result of the lowest order is
\begin{eqnarray}
\label{12}
- \frac{\vec{p}_1\vec{p}_2}{m_1 m_2 c^2\vec{k}^2}\rightarrow \frac{4\zeta(3)e^2}{3\pi \beta^3m_1 m_2 c^2}r_{12}^2(\vec{p}_1\vec{p}_2).
\end{eqnarray}

The Fourier transform for the next four terms can be performed using the substitution $\vec{k}\rightarrow -i\vec{\nabla}_1$. Acting by the gradient operator on the expression arising after angular integration, we find
\begin{eqnarray}
\label{13}
\frac{i\vec{\sigma}_1[\vec{k}\times\vec{p}_1]}{4m_1^2 c^2 \vec{k}^2} \rightarrow -\frac{2\zeta(3)e^2}{3\pi \beta^3 m_1^2 c^2}\left(\vec{\sigma}_1[\vec{r}_{12}\times\vec{p}_1]\right),
\nonumber
\\
-\frac{i\vec{\sigma}_2[\vec{k}\times\vec{p}_2]}{4m_2^2 c^2 \vec{k}^2} \rightarrow  \frac{2\zeta(3)e^2}{3\pi \beta^3 m_2^2 c^2}\left(\vec{\sigma}_2[\vec{r}_{12}\times\vec{p}_2]\right),
\\
\nonumber
-\frac{i\vec{\sigma}_1[\vec{k}\times\vec{p}_2]}{2m_1 m_2 c^2 \vec{k}^2}  \rightarrow \frac{4\zeta(3)e^2}{3\pi \beta^3 m_1 m_2 c^2}\left(\vec{\sigma}_1[\vec{r}_{12}\times\vec{p}_2]\right),
\\
\nonumber
\frac{i\vec{\sigma}_2[\vec{k}\times\vec{p}_1]}{2m_1 m_2 c^2 \vec{k}^2} \rightarrow -\frac{4\zeta(3)e^2}{3\pi \beta^3 m_1 m_2 c^2}\left(\vec{\sigma}_2[\vec{r}_{12}\times\vec{p}_1]\right).
\end{eqnarray}
Similar calculations for the last two contributions leads to
\begin{eqnarray}
\label{14}
\frac{(\vec{\sigma}_1\vec{k})(\vec{\sigma}_2\vec{k})}{4m_1 m_2 c^2\vec{k}^2}\rightarrow -\frac{4\zeta(5)e^2}{5\pi\beta^5 m_1 m_2 c^2}\times
\nonumber
\\
\left[r_{12}^2(\vec{\sigma}_1\vec{\sigma}_2)+2(\vec{r}_{12}\vec{\sigma}_1)(\vec{r}_{12}\vec{\sigma}_2\right]
\\
\nonumber
-\frac{\vec{\sigma}_1\vec{\sigma}_2}{4m_1 m_2 c^2}\rightarrow \frac{4\zeta(5)e^2}{\pi \beta^5 m_1 m_2 c^2}r_{12}^2(\vec{\sigma}_1\vec{\sigma}_2).
\end{eqnarray}
We emphasize that the replacement $\vec{k}\rightarrow -i\vec{\nabla}_1$ assumes first the action of the gradient operator and then going to the coincidence limit.

Finally, the total contribution in the lowest order in temperature can be obtained in the form:
\begin{eqnarray}
\label{15}
U(\vec{p}_1,\vec{p}_2,\vec{r}_{12}) = -\frac{4\zeta(3)e^2r_{12}^2}{3\pi\beta^3}  + \frac{8\zeta(3) e^2}{5\pi \beta^3 m_1 m_2 c^2}r_{12}^2(\vec{p}_1\vec{p}_2)
\nonumber
\\
+\frac{8\zeta(3) e^2}{15\pi \beta^3 m_1 m_2 c^2}\vec{r}_{12}(\vec{r}_{12}\vec{p}_2)\vec{p}_1
-\frac{2\zeta(3)e^2}{3\pi \beta^3 m_1^2 c^2}\left(\vec{\sigma}_1[\vec{r}_{12}\times\vec{p}_1]\right)
\nonumber
\\
\nonumber
+ \frac{2\zeta(3)e^2}{3\pi \beta^3 m_2^2 c^2}\left(\vec{\sigma}_2[\vec{r}_{12}\times\vec{p}_2]\right) + \frac{4\zeta(3)e^2}{3\pi \beta^3 m_1 m_2 c^2}\left(\vec{\sigma}_1[\vec{r}_{12}\times\vec{p}_2]\right) 
\\
-\frac{4\zeta(3)e^2}{3\pi \beta^3 m_1 m_2 c^2}\left(\vec{\sigma}_2[\vec{r}_{12}\times\vec{p}_1]\right).\qquad
\end{eqnarray}

\section{Hydrogen atom}

The expressions (\ref{4}), (\ref{7}) and (\ref{11})-(\ref{14}) allow significant reduction for the hydrogen atom, when the approximation of the infinite nucleus mass is considered. Then the thermal corrections of lowest order in temperature and $\alpha$ (the fine structure constant) arises from Eq. (\ref{15}) as
\begin{eqnarray}
\label{16}
U(\vec{p}_1,\vec{r}_{12}) = -\frac{4\zeta(3)e^2r_{12}^2}{3\pi\beta^3} 
-\frac{2\zeta(3)e^2}{3\pi \beta^3 m_1^2 c^2}\left(\vec{\sigma}_1[\vec{r}_{12}\times\vec{p}_1]\right).
\end{eqnarray}
To the latter the approximation of the point-like nucleus can be applied, when $\vec{r}_{12}$ can be replaced by the vector $\vec{r}$ representing the distance between the bound electron and the nucleus, $m_1$ is the electron mass, and we can take into account that $\vec{\sigma}_1 = \frac{1}{2}\vec{s}$ corresponds to the electron spin. Noticing that the operator $[\vec{r}\times\vec{p}]\equiv \vec{l}$ is the orbital momentum, one can find
\begin{eqnarray}
\label{17}
U(\vec{p},\vec{r}) = -\frac{4\zeta(3)Ze^2}{3\pi\beta^3} r^2
-\frac{4\zeta(3)Ze^2}{3\pi \beta^3 m^2 c^2}(\vec{s}\;\vec{l}),
\end{eqnarray}
where $Z$ denotes the nuclear charge.

The parametric estimation in relativistic units for the expression (\ref{17}) arises with $r\sim 1/(m\alpha Z)$, $\beta\sim m\alpha^2 (k_B T)$, $c=1$. Then the first term is proportional to $m\alpha^5 (k_B T)^3/Z$ and the second term to $ m \alpha^7 Z (k_B T)^3$. To get an estimate in atomic units, this results must be divided by a factor $m\alpha^2$. The evaluation of the $(\vec{s}\vec{l})$ operator can be done using the relation $\vec{j}^2=(\vec{l}+\vec{s})^2$, which results to the average value $\langle (\vec{s}\vec{l})\rangle = \frac{1}{2}[j(j+1)-l(l+1)-s(s+1)]$. Whereas the average value $r^2$ for the state $a$ in the hydrogen atom is $\frac{n_a^2}{2}(5n_a^2+1-3l_a(l_a+1))$, the energy shift taking into account the fine structure in the lowest order in atomic units is
\begin{eqnarray}
\label{18}
\Delta E_a = -\frac{2\zeta(3)}{3\pi\beta^3}\alpha^3 n_a^2[5n_a^2+1-3l_a(l_a+1)]
\\
\nonumber
 -\frac{2\zeta(3)}{3\pi\beta^3}\alpha^5[j_a(j_a+1)-l_a(l_a+1)-s_a(s_a+1)].
\end{eqnarray}

As a next step one can take into account the effect of the finite nuclear mass in the lowest order. For this, the nuclear momentum $\vec{p}_2$ is replaced by the $-\vec{p}_1$ in the center-of-mass system. Then in approximation of the point-like nucleus, the result is
\begin{eqnarray}
\label{19}
U(\vec{p},\vec{r}) = -\frac{4\zeta(3)Ze^2}{3\pi\beta^3} r^2 + \frac{4\zeta(5)Ze^2}{5\pi \beta^5} r^4 + \frac{2\zeta(5)Ze^2}{\pi\beta^5 m^2 c^2} r^2 \qquad
\\
\nonumber
+
 \frac{2\zeta(5)Ze^2}{\pi\beta^5 M^2 c^2} r^2 -\frac{8\zeta(3)Ze^2}{5\pi\beta^3 m M c^2} r^2 p^2 - \frac{8\zeta(3)Ze^2}{15\pi\beta^3 m M c^2} \vec{r}(\vec{r}\vec{p})\vec{p} 
\\
\nonumber
-
\frac{4\zeta(3)Ze^2}{3\pi \beta^3 m^2 c^2}\left[1+\frac{2m}{M}\right] (\vec{s}\vec{l}) -\frac{8\zeta(3)Ze^2}{3\pi \beta^3 m M c^2}\left[1+\frac{m}{2M}\right] (\vec{I}_p\vec{l}) 
\\
\nonumber
+
\frac{64\zeta(5)Ze^2}{5\pi\beta^5 m M c^2} r^2(\vec{s}\vec{I}_p)-\frac{32\zeta(5)Ze^2}{5\pi \beta^5 m M c^2} (\vec{r}\vec{s})(\vec{r}\vec{I}_p),
\end{eqnarray}
where $\vec{I}_p$ denotes the nuclear spin operator and $M\equiv m_2$ is the nuclear mass. The meaning of these terms can be found in textbooks on quantum electrodynamics, see also \cite{Daza_2012}, with the additional notation 'thermal'.

In a hydrogen atom at room temperature, corrections (\ref{19}) can be neglected except for corrections (\ref{18}). 
For clarity, the values of the thermal corrections Eq. (\ref{19}) are shown in Table~\ref{tab:1} for specific low-lying states at room temperature.
\begin{widetext}
\begin{center}
\begin{table}[ht!]
\caption{The values of thermal corrections Eq. (\ref{19}) in Hz. The first column gives the thermal correction (Th. corr.). The following columns show the values obtained for the specific states. The dependence on the nuclear charge $Z$ is left to determine the behavior of corrections for hydrogen-like ions.}
\label{tab:1}
\begin{tabular}{ c | c | c | c | c}
\hline
\hline
Th. corr. & $1s$ & $2s$ & $2p_{1/2}$ & $2p_{3/2}$\\
\hline
\noalign{\smallskip}
$-\frac{4\zeta(3)Ze^2}{3\pi\beta^3} r^2$ & $-3.35\frac{1}{Z}$ & $46.98\frac{1}{Z}$ & $33.55\frac{1}{Z}$ & $33.55\frac{1}{Z}$\\

\hline
\noalign{\smallskip}
$\frac{4\zeta(5)Ze^2}{5\pi \beta^5} r^4$ & $6.26\cdot 10^{-10}\frac{1}{Z^3}$  & $ 8.01\cdot 10^{-8}\frac{1}{Z^3}$ & $4.67\cdot 10^{-8}\frac{1}{Z^3}$ & $4.67\cdot 10^{-8}\frac{1}{Z^3}$\\

\hline
\noalign{\smallskip}
$\frac{2\zeta(5)Ze^2}{\pi\beta^5 c^2} r^2\left[\frac{1}{m^2}+\frac{1}{M^2}\right]$ & $1.11\cdot 10^{-14}\frac{1}{Z}$ & $1.56\cdot 10^{-13}\frac{1}{Z}$ &  $1.11\cdot 10^{-13}\frac{1}{Z}$ & $1.11\cdot 10^{-13}\frac{1}{Z}$\\

\hline
\noalign{\smallskip}
$ -\frac{8\zeta(3)Ze^2}{5\pi\beta^3 m M c^2} r^2 p^2$ & $-1.18\cdot 10^{-7}Z$ & $-4.12\cdot 10^{-7}Z$ & $-2.94793\cdot 10^{-7}Z$ & $-2.94789\cdot 10^{-7}Z$\\

\hline
\noalign{\smallskip}
$- \frac{8\zeta(3)Ze^2}{15\pi\beta^3 m M c^2} \vec{r}(\vec{r}\vec{p})\vec{p}$ & $-3.92\cdot 10^{-8}Z$ & $-1.37\cdot 10^{-7}Z$ & $-7.23135\cdot 10^{-8}Z$ & $-7.23122\cdot 10^{-8}Z$\\

\hline
\noalign{\smallskip}
$-\frac{4\zeta(3)Ze^2}{3\pi \beta^3 m^2 c^2}\left[1+\frac{2m}{M}\right] (\vec{s}\vec{l})$ & $0$  & $0$ & $1.19\cdot 10^{-4}Z$ & $-5.96\cdot 10^{-5}Z$\\

\hline
\noalign{\smallskip}
\multirow{2}{*}{$-\frac{8\zeta(3)Ze^2}{3\pi \beta^3 m M c^2}\left[1+\frac{m}{2M}\right] (\vec{I}_p\vec{l})$} & $0$ ($F=0$) & $0$ ($F=0$) & $6.49\cdot 10^{-8}Z$ ($F=0$) & $1.62\cdot 10^{-7}Z$ ($F=0$)\\
 & $0$ ($F=1$) & $0$ ($F=1$) & $-6.49\cdot 10^{-8}Z$ ($F=1$) & $-1.62\cdot 10^{-7}Z$ ($F=1$)\\
 
   \hline
  \noalign{\smallskip}  
\multirow{2}{*}{$\frac{64\zeta(5)Ze^2}{5\pi\beta^5 m M c^2} r^2(\vec{s}\vec{I}_p)$} & $-2.91\cdot 10^{-17}\frac{1}{Z}$ ($F=0$)  & $-4.07\cdot 10^{-16}\frac{1}{Z}$ ($F=0$) & $-9.68\cdot 10^{-17}\frac{1}{Z}$ ($F=0$) & $-5.38\cdot 10^{-17}\frac{1}{Z}$ ($F=1$)\\
& $3.23\cdot 10^{-18}\frac{1}{Z}$ ($F=1$) & $4.52\cdot 10^{-17}\frac{1}{Z}$ ($F=1$) & $1.08\cdot 10^{-17}\frac{1}{Z}$ ($F=1$) & $9.68\cdot 10^{-17}\frac{1}{Z}$ ($F=2$)\\

\hline
\noalign{\smallskip}
\multirow{2}{*}{$-\frac{32\zeta(5)Ze^2}{5\pi \beta^5 m M c^2} (\vec{r}\vec{s})(\vec{r}\vec{I}_p)$} & $4.84\cdot 10^{-18}\frac{1}{Z}$ ($F=0$)  & $6.78\cdot 10^{-17}\frac{1}{Z}$ ($F=0$) & $4.84\cdot 10^{-17}\frac{1}{Z}$ $(F=0)$ & $1.61\cdot 10^{-17}\frac{1}{Z}$ $(F=1)$ \\
&$-1.61\cdot 10^{-18}\frac{1}{Z}$ ($F=1$) & $-2.26\cdot 10^{-17}\frac{1}{Z}$ ($F=1$) & $-1.61\cdot 10^{-17}\frac{1}{Z}$ ($F=1$) & $-9.68\cdot 10^{-18}\frac{1}{Z}$ $(F=2)$\\

\hline
\hline
\end{tabular}
\end{table}
\end{center}
\end{widetext}

In particular, from Table~\ref{tab:1} it follows that the relativistic thermal corrections in the hydrogen atom are negligible at room temperature. Nevertheless, there are corrections that increase with increasing nuclear charge $Z$. The most interesting in this sense is the correction corresponding to the thermal shift of the fine sublevel and proportional to $\vec{s}\vec{l}$. Note also that in the nonrelativistic limit the thermal correction for the motion of the nucleus can be found in the lowest order, see \cite{Bethe}, by replacing the electron mass with the reduced one. Then, for the results listed in the first row of Table~\ref{tab:1}, one can find that the thermal correction on the finite mass of the nucleus is about $10^{-2}$ Hz (multiplication factor is $1/1836.15$), representing, thus, a leading order correction with respect to relativistic corrections. Having made rough estimates for highly charged hydrogen-like ions, it can be found that these corrections still go beyond the accuracy of laboratory experiments. However, assuming the astrophysical applications, one can expect a cubic increase of the thermal corrections with a rise in temperature.

\section{Helium atom}

Operator Eq. (\ref{15}) admits the evaluation of the relativistic thermal corrections for the helium atom. For this, it is convenient to generalize the formula (\ref{15}) to the case of an arbitrary number of electron-electron and electron-nuclear interactions:
\begin{eqnarray}
\label{20}
U = -\frac{4\zeta(3) e^2}{3\pi\beta^3}\sum\limits^{N+1}_{i<j}Z_iZ_j r_{ij}^2  
\nonumber
\\
+ \frac{8\zeta(3) e^2}{15\pi \beta^3 c^2}\sum\limits^{N+1}_{i<j}\frac{Z_iZ_j}{m_im_j}\left[3r_{ij}^2(\vec{p}_i\vec{p}_j)+\vec{r}_{ij}(\vec{r}_{ij}\vec{p}_j)\vec{p}_i\right]
\\
\nonumber
-\frac{2\zeta(3)e^2}{3\pi \beta^3 c^2}\sum\limits^{N+1}_{\substack{i,j=1\\i\neq j}}\frac{Z_iZ_j}{m_i^2}\left(\vec{\sigma}_i[\vec{r}_{ij}\times\vec{p}_i]\right) 
\nonumber
\\
\nonumber
 + \frac{4\zeta(3)e^2}{3\pi \beta^3 c^2}\sum\limits^{N+1}_{\substack{i,j=1\\i\neq j}}\frac{Z_iZ_j}{m_im_j}\left(\vec{\sigma}_i[\vec{r}_{ij}\times\vec{p}_j]\right) ,
\end{eqnarray}
where $Z_i$ represents the corresponding charge of the particle and summation involves all particles including the nucleus. In case of two-electron ion (i.e. when $N=2$ in Eq. (\ref{20})) with nuclear charge $Z_{3}=Z$ and electron charges $Z_{1}=Z_{2}=-1$ the corresponding electron and nuclear masses are $m_{1}=m_{2}=m$ and $m_{3}=M$, respectively. 

In this case two types of interaction should be described. The first corresponds to the electron-nucleus and the other is given by the interelectron thermal interaction. Then the lowest order thermal correction, see \cite{SZA-2020}, can be written as
\begin{eqnarray}
\label{21}
\langle U^{(1)} \rangle = -\frac{4\zeta(3)Ze^2}{3\pi\beta^3}\langle r_1^2+r_2^2-\frac{1}{Z}r_{12}^2\rangle,
\end{eqnarray}
where $r_i\equiv r_{i3}$ is the distance between the corresponding electron and the nucleus, $r_{12}$ represents the modulus of the radius vector between the electrons and $\langle\dots\rangle$ denotes the averaging on the atomic state. The calculated expectation values of the  $r_1^2$, $r_{12}^2$ operators and the corresponding energy shifts are collected in Table~\ref{tab:2}.
\begin{table}[ht]
\begin{center}
\caption{Expectation values of $r_1^2$, $r_{12}^2$ operators (in a.u.) and the corresponding energy shift $\Delta E^\beta \equiv \langle U^{(1)} \rangle$, see Eq. (\ref{21}), at room temperature ($T=300$K) in Hz for the He($M = \infty$) atom.
}
\label{tab:2}
\begin{tabular}{ c |  c |  c |  c |  c}
\hline
\hline
\noalign{\smallskip}
State & $\langle r_{12}^2 \rangle $ in a.u. & $\langle r_1^2 \rangle$ in a.u. & $\Delta E^\beta $ in a.u.  & $\Delta E^\beta$ in Hz \\
\hline
\noalign{\smallskip}
$ 1^1S $ & $ 2.516439313 $  & $ 1.193482995 $ & $ -3.83766\cdot 10^{-16} $ & $ -2.52505 $\\
\hline
\noalign{\smallskip}
$ 2^1S $ & $ 32.30238038 $  & $ 16.08923325 $ & $ -5.44916\cdot 10^{-15} $ & $ -35.8537 $\\
$ 2^3S $ & $ 23.04619748 $  & $ 11.46432162 $ & $ -3.8778\cdot 10^{-15} $ & $ -25.5147 $\\
\hline
\noalign{\smallskip}
$ 2^1P $ & $ 31.59851603 $  & $ 15.76565497 $ & $ -5.34878\cdot 10^{-15} $ & $ -35.1933 $\\
$ 2^3P $ & $ 26.64279322 $  & $ 13.21174046 $ & $ -4.45461\cdot 10^{-15} $ & $ -29.3099 $\\
\hline
\noalign{\smallskip}
$ 3^1S $ & $ 171.8387553 $  & $ 85.89015822 $ & $ -2.91921\cdot 10^{-14} $ & $ -192.075 $\\
$ 3^3S $ & $ 137.4750832 $  & $ 68.70840131 $ & $ -2.33505\cdot 10^{-14} $ & $ -153.638 $\\
\hline
\noalign{\smallskip}
$ 3^1P $ & $ 183.7866266 $  & $ 91.87290715 $ & $ -3.12292\cdot 10^{-14} $ & $ -205.478 $\\
$ 3^3P $ & $ 164.3028806 $  & $ 82.10989293 $ & $ -2.79026\cdot 10^{-14} $ & $ -183.59 $\\
\hline
\noalign{\smallskip}
$ 3^1D $ & $ 126.4161413 $  & $ 63.17681865 $ & $ -2.1469\cdot 10^{-14} $ & $ -141.259 $\\
$ 3^3D $ & $ 126.2834766 $  & $ 63.11075331 $ & $ -2.14467\cdot 10^{-14} $ & $ -141.112 $\\
\hline
\noalign{\smallskip}
$ 4^1S $ & $ 562.8623853 $ & $ 281.4144002 $ & $ -9.56731\cdot 10^{-14} $ & $ -629.499 $\\
$ 5^1S $ & $ 1406.742766 $ & $ 703.3605371 $ & $ -2.39134\cdot 10^{-13} $ & $ -1573.42 $\\
\hline
\hline
\end{tabular}
\end{center}
\end{table}

The thermal correction on the motion of nucleus can be found from Table~\ref{tab:2} by dividing the results by $7294.3$ (alpha particle-electron mass ratio). For the numerical calculations in the two-electron atom, we use trial wave functions with quasirandom nonlinear parameters developed in \cite{Korobov_1999,Korobov_2000}. The calculated expectation values of the $r_1^2$, $r_{12}^2$ operators are in excellent agreement with \cite{Drakebook,Frolov-1998}.

According to the results of \cite{SZA-2020} the found thermal correction Eq. (\ref{21}), see Table~\ref{tab:2}, reaches the level of experimental accuracy \cite{KS-Hessel} at room temperature. Thermal corrections of the next order in $\alpha$, of the same order of temperature, and in the approximation of infinite nuclear mass are
\begin{gather}
\label{22}
\langle U^{(2)} \rangle =
\frac{8\zeta(3)e^2}{15\pi\beta^3m^2c^2}\langle\left(3r_{12}^2(\vec{p}_1\vec{p}_2)+\vec{r}_{12}(\vec{r}_{12}\vec{p}_2)\vec{p}_1\right)\rangle
\\
\nonumber
-\frac{4\zeta(3)Ze^2}{3\pi\beta^3 m^2c^2}\langle (\vec{s}_1 \vec{l}_1) + (\vec{s}_2 \vec{l}_2)   \qquad
\\
\nonumber
+
\frac{1}{Z}
\left(
[\vec{r}_{12}\times\vec{p}_1](\vec{s}_1+2\vec{s}_2)-[\vec{r}_{12}\times\vec{p}_2](\vec{s}_1+2\vec{s}_2)
\right)
\rangle
\end{gather}
The numerical values of these operators and energy shift are collected in Table~\ref{tab:3}.

\begin{table}[ht]
\begin{center}
\caption{Expectation values of $U^{(2)}_A=(\vec{s}_1 \vec{l}_1) + (\vec{s}_2 \vec{l}_2)$, $U^{(2)}_B = \frac{1}{Z}
\left(
[\vec{r}_{12}\times\vec{p}_1](\vec{s}_1+2\vec{s}_2)-[\vec{r}_{12}\times\vec{p}_2](\vec{s}_1+2\vec{s}_2)
\right)$  and $U^{(2)}_C = 3r_{12}^2(\vec{p}_1\vec{p}_2)+\vec{r}_{12}(\vec{r}_{12}\vec{p}_2)\vec{p}_1$ operators (in a.u.) and the corresponding total energy shift $\Delta E^\beta \equiv \langle U^{(2)} \rangle $, see Eq. (\ref{22}), at room temperature ($T=300$K) in Hz for the He($M = \infty$) atom.
}
\label{tab:3}
\begin{tabular}{ c |  c |  c |  c |  c}
\hline
\hline
\noalign{\smallskip}
State & $\langle U^{(2)}_A \rangle$ in a.u. & $\langle U^{(2)}_B \rangle$ in a.u. & $\langle U^{(2)}_C \rangle$ in a.u. & $\Delta E^\beta$ in Hz\\
\hline
$ 1^1S_0 $ & $0$  & $0$  & $7.69941$  & $1.834\cdot 10^{-4}$\\
\hline
$ 2^1S_0 $ & $0$  & $0$ & $7.19352$  & $1.714\cdot 10^{-4}$\\
$ 2^3S_1 $ & $0$  & $0$ & $7.99171$  & $1.904\cdot 10^{-4}$\\
\hline
$ 2^1P_1 $ & $0$  & $0$ & $4.77589$  & $1.139\cdot 10^{-4}$\\
$2^3P_0$ & $-1.$ & $1.69266$ & $3.58270$  & $2.845\cdot 10^{-6}$\\
$2^3P_1$ & $-0.5$ & $0.846328$ & $3.58270$  & $4.410\cdot 10^{-5}$\\
$2^3P_2$ & $0.5$ & $-0.846328$ & $3.58270$  & $1.266\cdot 10^{-4}$\\
\hline
$ 3^1S_0 $ & $0$  & $0$ & $7.40854$  & $1.765\cdot 10^{-4}$\\
$ 3^3S_1 $ & $0$  & $0$ & $7.59006$  & $1.808\cdot 10^{-4}$\\
\hline
$ 3^1P_1 $ & $0$  & $0$ & $4.45017$  & $1.060\cdot 10^{-4}$\\
$ 3^3P_0 $ & $-1.$  & $1.55232$ & $4.14054$  & $3.285\cdot 10^{-5}$\\
$ 3^3P_1 $ & $-0.5$  & $0.776158$ & $4.14054$  & $6.575\cdot 10^{-4}$\\
$ 3^3P_2 $ & $0.5$  & $-0.776158$ & $4.14054$  & $1.315\cdot 10^{-4}$\\

\hline
\hline
\end{tabular}
\end{center}
\end{table}

The total contribution of thermal corrections Eq. (\ref{22}), listed in Table~\ref{tab:3} for different states in helium, does not exceed several parts of $10^{-4}$. With that, the thermal shift arising due to the finite mass of the nucleus (dividing the results of Table~\ref{tab:2} by the nuclear mass) may turn out to be significant for highly excited states. For example, for the $3^3P$ state, it is about $0.028$ Hz, whereas for the $5^1S$ it reaches $0.21$ Hz. The most problem in this context arises with the method of calculating the binding energies of highly excited states in the helium atom. Moreover, in \cite{SZA-vertex} it can be found that the calculations of the lowest order thermal shift for highly excited states must be performed with a closed form of thermal potential, Eq. (\ref{4}). Still, the calculations presented here can serve to pick out the trend of thermal shifts (\ref{21}) and (\ref{22}) in this case. First, as can be found analytically, the energy shift corresponding to the thermal spin-orbit interaction does not depend on the principal quantum number (almost the same). Second, the energy shift, Eq. (\ref{21}), has the order of a thermal Stark shift and always remains negative as opposed to a Stark shift, see \cite{farley}.

\section{Conclusions and discussion}

In this article, we examined relativistic corrections due to thermal interaction. As was found in \cite{S-TQED}, applying the rigorous QED theory to the description of the interaction of two charges placed in a heat bath reveals effects that do not arise in the framework of the quantum mechanical approach. As a consequence, the thermal potential Eq. (\ref{4}) was obtained in \cite{S-TQED}, where the leading-order radiative corrections were also described. 

To complete the QED description of thermal effects, it is necessary to take into account the relativistic corrections to the interaction potential. The procedure is greatly simplified by virtue of the form of thermal photon propagator Eq. (\ref{2}). First of all, it admits a simple introduction of thermal gauges, see \cite{S-TQED}. Then, by choosing the thermal Coulomb gauge, the relativistic corrections to the thermal interaction can be easily written by analogy with the formula (\ref{5}) in the momentum space, see \cite{Berest}. The difference from the ordinary (zero vacuum) case is the presence of contour integration and the Planck's distribution function. Then, by performing the sequential calculations, the expression (\ref{15}) can be found. In a hydrogen atom, it reduces to Eq. (\ref{19}) and then can be generalized to the expression (\ref{20}) in the case of an arbitrary number of charges.

The expression (\ref{20}) allows parametric estimation in the fine structure constant. For the first term, it reads $\alpha^3/Z\cdot (k_B T)^3$ in atomic units. Then the leading-order relativistic corrections are $\alpha^2$ times less (as usual) and correspond to the spin-orbit thermal interaction. The found estimate by the nuclear charge number $Z$ shows a decrease of thermal correction Eq. (\ref{17}) for the hydrogen-like ions. The dependence on the nuclear charge $Z$ is left in Table~\ref{tab:1} to demonstrate inference. In the case, $Z=1$, one of the possible applications of the presented theory is the study of the proton form factor effects involving the spin-spin and spin-orbit thermal interaction \cite{Daza_2012}.
Another conclusion following from the expression (\ref{19}) is that the thermal relativistic corrections corresponding to the spin-orbit interaction do not depend on the principal quantum number and, more interestingly, are proportional to the nuclear charge number $Z$. The results listed in Table~\ref{tab:1}, however, demonstrate that these corrections are still outside the experimental accuracy for arbitrary $Z$. 

Nonetheless, the dependence only on angular momenta allows an approximate estimate of the correction Eq. (\ref{18}) for fine sublevels in other atoms. Consider, for example, a precision measurement of atomic isotope shifts in ${}^{86}$Sr$^+$ and ${}^{88}$Sr$^+$ based on $5S_{1/2}\leftrightarrow 4D_{5/2}$ transition \cite{Sr-2019}. The thermal shift proportional to $(\vec{s}\vec{l})$ for the $4d_{5/2}$ state in hydrogen is $-0.000119$ Hz at room temperature and an approximate estimate in Sr$^+$ is $-0.0045$ Hz, while the declared error is about $0.009$ Hz. Another example can be given for the neutral Sr atom. Performing the same estimates using the results in Table~\ref{tab:3}, a thermal correction of the order of $-0.005$ Hz can be found, resulting in a relative value of about $-1.2\times 10^{-17}$ (which exceeds the dc-Stark effect on the order) for the measured frequency between the ground state ${}^1S_0$ and the metastable state ${}^3P_0$ \cite{Sr-clock}. Especially, in \cite{Sr-clock} it was assumed that the planned controlling the BBR shift would increase the accuracy to $1\times 10^{-17}$ or below in a near future, making such thermal corrections potentially important. 

Moreover, since the thermal Stark shift can change sign with a varying in temperature, it was recently predicted in \cite{Martin-Safronova} that the rubidium atomic clock could be operated with zero net BBR shift at the temperature of about $495$ K. In turn, the thermal corrections Eq. (\ref{20}) would increase by about $4.6$ times when raised to this temperature. Despite such a rough analysis, this estimate discloses the need for a separate description of thermal effects in many-electron atomic systems. It should be noted that in this case the dominant contribution $-\frac{4\zeta(3)Ze^2}{3\pi\beta^3} r^2$ was not taken into account, which can only be calculated using specialized numerical methods.

Along with that, one should separately focus on the tendency of recent years towards the search for 'new physics' and verification of fundamental interactions within the framework of atomic physics. The achieved experimental accuracy at the level of $10^{-18}$ of operating atomic clocks with single atomic systems \cite{AtCl-Sr,AtCl-2016} makes them the most accurate tool for searching for dark matter setting constraints on its mass \cite{DM-atcl}. The study of new physics using the Rydberg states of atomic hydrogen was proposed in \cite{NewPhys-RydbergH}. All spectroscopic experiments with this level of accuracy require stabilization of the temperature environment and the corresponding consideration of the thermal Stark shift. The results obtained in this paper show the need to take into account additional thermal effects in such an analysis. In contrast to the traditional relativistic QM description of the thermal interaction determined by the multipolar method, \cite{Porsev,Sahoo2016} or, for example, the study of the attractive force induced by black body radiation between the atom and the heated cavity \cite{AttractiveForce}, the description given in this work reveals the fundamental nature of thermal interactions of another type. In addition, astrophysical prolongation of the discovered effects is evident, when the temperature can reach much higher values.

\section*{Acknowledgements}
This work was supported by Russian Foundation for Basic Research (grant 20-02-00111). 

\bibliography{mybibfile}

\appendix
\renewcommand{\theequation}{A\arabic{equation}}
\setcounter{equation}{0}
\section*{Appendix: reduction of matrix elements}

In this Appendix a most general formulas for the reduction of matrix elements for operators $\vec{s}_i [\vec{r}_i \times \vec{p}_i]$, $(\vec{s}_i+2\vec{s}_{j})[\vec{r}_{ij}\times\vec{p}_j]$ and $3r_{ij}^2(\vec{p}_i\vec{p}_j)+\vec{r}_{ij}(\vec{r}_{ij}\vec{p}_j)\vec{p}_i$ in the $LSJM$ coupling scheme ($L$ is the total orbital momentum, $S$ is the total spin, $J,M$ are the total angular momentum and its projection) are given. Since the spin operator and orbital momentum operator act on different subsystems, the expectation value of $\vec{s}_i \vec{l}_i$ operator can be written, see \cite{Varsh}, as
\begin{eqnarray}
\langle n'L'S'J'M'  |\vec{s}_i [\vec{r}_i \times \vec{p}_i]  | nLSJM\rangle = \delta_{J'J}\delta_{M'M}\times \qquad
\\
\nonumber
(-1)^{J+L+S'}
\begin{Bmatrix}
J & S' & L'\\
1 & L & S
\end{Bmatrix}
\langle n'L' ||r_i \times p_i || nL\rangle
\langle S' ||s_{i} || S\rangle
,
\end{eqnarray}
where
\begin{eqnarray}
\langle S' ||s_{1(2)} || S\rangle = (-1)^{S(S')}\sqrt{(2S'+1)(2S+1)}
\\
\nonumber
\times
\begin{Bmatrix}
1/2 & S' & 1/2\\
S & 1/2 & 1
\end{Bmatrix}
\sqrt{3/2}
\end{eqnarray}
and $ \langle n'L'||\dots || nL\rangle $ is the reduced matrix element. 

Similar equations can be written for expectation values of operator $ (\vec{s}_i+2\vec{s}_{j}) $
\begin{eqnarray}
\langle n'L'S'J'M'  |(\vec{s}_i+2\vec{s}_{j})[\vec{r}_{ij}\times\vec{p}_j]  | nLSJM\rangle 
\\
\nonumber
= \delta_{J'J}\delta_{M'M}
(-1)^{J+L+S'}
\begin{Bmatrix}
J & S' & L'\\
1 & L & S
\end{Bmatrix}
\\
\nonumber
\times
\langle n'L' ||r_{ij} \times p_j || nL\rangle
\langle S' ||s_{i}+2s_{j} || S\rangle
.
\end{eqnarray}

The latter operator $ 3r_{ij}^2(\vec{p}_i\vec{p}_j)+\vec{r}_{ij}(\vec{r}_{ij}\vec{p}_j)\vec{p}_i $ is scalar and do not acts on spin variables. Therefore, 
\begin{eqnarray}
\langle n'L'S'J'M'  | 3r_{ij}^2(\vec{p}_i\vec{p}_j)+\vec{r}_{ij}(\vec{r}_{ij}\vec{p}_j)\vec{p}_i | nLSJM\rangle
\\\nonumber
=\delta_{S'S}(-1)^{J+L'+S}\sqrt{2J+1}C^{J'M'}_{JM00}
\begin{Bmatrix}
L & S & J\\
J' & 0 & L'
\end{Bmatrix}
\\
\nonumber
\times
\langle n'L' || 3r_{ij}^2(p_ip_j)+r_{ij}(r_{ij}p_j)p_i || nL\rangle
.
\end{eqnarray}
Here $C^{J'M'}_{JM00}=\delta_{J'J}\delta_{M'M}$ is the Clebsch-Gordan coefficient and $\begin{Bmatrix}
j_1 & j_2 & j_3\\
j_4 & j_5 & j_6
\end{Bmatrix}$ gives the values of the Racah 6-j symbol.
\end{document}